\documentclass[preprint,12pt, a4paper]{elsarticle}

\usepackage{graphicx}
\usepackage{balance}
\usepackage{pgfplots}
\pgfplotsset{compat=1.18} 
\usepackage{hyperref}
\usepackage{graphicx}
\usepackage{subcaption}
\usepackage{makecell}
\usepackage{multicol}
\usepackage{pifont}
\usepackage{multirow}
\usepackage{hhline}
\usepackage{booktabs}
\usepackage{tcolorbox}
\usepackage{listings}
\usepackage{url}
\def\BibTeX{{\rm B\kern-.05em{\sc i\kern-.025em b}\kern-.08em
    T\kern-.1667em\lower.7ex\hbox{E}\kern-.125emX}}
\AtBeginDocument{\definecolor{tmlcncolor}{cmyk}{0.93,0.59,0.15,0.02}\definecolor{NavyBlue}{RGB}{0,86,125}}

\newcommand{\cmark}{\ding{51}}%
\newcommand{\xmark}{\ding{55}}%

\begin{document}

\begin{frontmatter}
\author[udc]{David Soler\texorpdfstring{\corref{cor1}}{*}}
\ead{david.soler@udc.es}
\cortext[cor1]{Corresponding author}

\title{Analysis of Efficiency of the Messaging Layer Security protocol in Experimental Settings}

\author[udc]{Carlos Dafonte}
\author[uvigo]{Manuel Fernández-Veiga}
\author[uvigo]{Ana Fernández Vilas}
\author[udc]{Francisco J. Nóvoa}


\begin{abstract}
Messaging Layer Security (MLS) and its underlying Continuous Group Key Agreement (CGKA) protocol allows a group of users to share a cryptographic secret in a dynamic manner, such that the secret is modified in member insertions and deletions. One of the most relevant contributions of MLS is its efficiency, as its communication cost scales logarithmically with the number of members. However, this claim has only been analysed in theoretical models and thus it is unclear how efficient MLS is in real-world scenarios. Furthermore, practical considerations such as the chosen paradigm and the evolution of the group can also influence the performance of an MLS group. In this work we analyse MLS from an empirical viewpoint: we provide real-world measurements for metrics such as commit generation and processing times and message sizes under different conditions. In order to obtain these results we have developed a highly configurable environment for empirical evaluations of MLS through the emulation of MLS clients. Among other findings, our results show that computation costs scale linearly in practical settings even in the best-case scenario.
\end{abstract}



\end{frontmatter}

\section{Introduction} 
\label{sec:introduction}

Messaging Layer Security (MLS)~\cite{mls} is a recent communications standard for establishing secure messaging groups between a set of users. It is mainly composed of a cryptographic scheme named Continuous Group Key Agreement (CGKA), whose objective is to distribute a shared secret between members of the group. One of the most relevant characteristics of CGKA protocols is its flexibility in group composition: the scheme allows insertions and deletions of members while ensuring the security properties of Forward Secrecy (FS) and Post-Compromise Security (PCS). 

In addition to these security properties CGKA schemes also have a significant focus in maintaining efficiency as the group increases in size. Indeed, MLS employs a version of TreeKEM \cite{treekem} as its underlying CGKA protocol, which structures the group as a binary tree in order to achieve logarithmic complexity in relation to group size. The efficiency of CGKA protocols has been widely analysed in the literature \cite{cost_1, cost_2, bounds} and multiple CGKA variants that attempt to increase efficiency under certain scenarios have been proposed \cite{saik, cmpke, decaf, overlap}. 

However, most contributions to the CGKA literature only address the theoretical aspects of these protocols. Although some works perform empirical measurements of their CGKA variants \cite{a_cgka, dec_ack, cost_tree}, they are often limited to small groups and specific scenarios. Furthermore, MLS introduces new elements to CGKA groups whose impact has not been studied in the literature. In particular, MLS defines a method for \textit{External Joins}, through which users can enter a group without requiring invitation. The KeyPackage and GroupInfo messages are required to add new members to the group, and they need to be stored and made available to users. The \textit{propose-and-commit} paradigm requires the generation, distribution and processing of proposal messages that represent an additional overhead.


In this work we address those issues by performing an empirical evaluation of the performance of MLS under multiple different scenarios. We analyse how measurements such as commit generation, processing and distribution times or message size are affected by different experimental conditions: the size of the group, the chosen paradigm, or the evolution of the group state. 

To that end we have developed a testbed for experimental evaluations of MLS. Our implementation consists of a emulated MLS client that autonomously joins groups and periodically publishes messages and updates, as well as two different Delivery Services to distribute MLS messages between the clients. The testbed is highly configurable and allows for the deployment of an arbitrary number of emulated clients. We publish our implementation as open-source so that it is available for researchers to perform their own analysis or even create new scenarios.

In summary, our work presents the following contributions:
\begin{enumerate}
    \item A testbed for experimental evaluations of MLS. We provide a Rust implementation\footnote{Available at \url{https://github.com/SDABIS/mls_experimental_analysis}} of a emulated MLS client that can be configured in detail to adapt their behaviour. When executed in parallel, these clients will create MLS groups, add or remove other clients and exchange application messages, according to their parameters.
    \item An empirical analysis of the performance of MLS. We define various \textit{experiments} in which the emulated clients are configured to recreate MLS groups with different configurations. The experiments performed study the impact of MLS functionalities like External Joins and the \textit{propose-and-commit}, as well as the degradation of the group's state by blanked intermediate nodes. We measure the computational cost of generating and processing updates, the size of exchanged messages and the time required to deliver them to all members. Among other findings, our results show the relevance of practical considerations such as the chosen paradigm and persistent storage. Furthermore, we identify that the expected logarithmic complexity of MLS does not hold in practice even in a best-case scenario due to the high cost of managing the group state, whose size increases linearly as the number of members grow.
\end{enumerate}

The rest of the document is organised as follows: Section~\ref{sec:literature} will review related works to provide the context to our contribution. Section~\ref{sec:background} will introduce to the reader the concepts of CGKA and MLS that are relevant for our analysis. In Section~\ref{sec:env} we describe our testbed and the experiments to be performed. Section~\ref{sec:impl} introduces our implementation of the emulated MLS clients and the Delivery Service and in Section~\ref{sec:result} we present our measurements of said environment. In Section \ref{sec:discussion} we discuss our most relevant findings and overall implications of our results. Finally, Section~\ref{sec:conclusion} will conclude this document and discuss future improvements.

\section{Related Work}
\label{sec:literature}

CGKA protocols have been thoroughly defined in the literature~\cite{cgka_analysis, itk, pass_continuous_2020}, exploring the security properties of Forward Security (FS) and Post-Compromise Security (PCS)~\cite{cgka_analysis, worst-case}. Since one of the main objectives of CGKA protocols is to provide efficient  scaling with the number of users~\cite{bounds}, they usually employ binary trees to represent the state of the group~\cite{art, treekem, ttkem}. The most popular instantiation of a CGKA protocol is the Messaging Layer Security (MLS), which has recently been standardised as RFC 9420 \cite{mls}. 


The efficiency of CGKA protocols has mostly been discussed in a theoretical framework, without presenting experimental results. The authors of \cite{saik} discuss the bandwidth required to distribute messages to all members of the CGKA group, and propose introducing a server to more efficiently provide each member with the information they require. A similar approach is taken in \cite{cmpke}, which also optimises the protocol for post-quantum algorithms for public key encryption. The impact of the tree's state in efficiency is taken into account. Other works further develop this analysis to estimate the \textit{communication cost} (i.e., the amount of messages required) of healing the group after a compromise. Communication cost is estimated both for a generic CGKA protocol \cite{cost_1, cost_2, bounds} and for specific schemes \cite{cocoa, qtk, cgka_fa}. The authors of \cite{cost_tree} also analyse the communication cost of the TreeKEM protocol by focusing on the shape of the ratchet tree.

The authors of \cite{discreet_original} propose a Delivery Service that employs Reliable Broadcast and consensus mechanisms to distribute proposals and commits, respectively. They later presented an implementation \cite{discreet}, but without experimental measurements. The CGKA variant introduced in \cite{decaf} is particularly suited for its implementation using a Blockchain as a decentralised Delivery Service, although the work is mainly theoretical and no experimental environment is discussed. Conversely, the authors of \cite{art_bc} do implement a CGKA environment using a blockchain for IoT devices, but employ ART \cite{art}, an outdated version of TreeKEM and thus not compatible with MLS. In \cite{dec_ack} an alternative CGKA protocol that does not employ binary trees is presented and an implementation is provided for a simplified execution environment. 

Experimental analysis of CGKA implementations is limited to works that present an specific CGKA variant such as \cite{a_cgka, dec_ack, treesync}, although their experimental settings are limited: they only consider groups with few members ---up to 128--- and do not model user behaviour. In both cases their measurements show how commit generation times increase linearly as the number of users grow. 

\section{Background} 
\label{sec:background}

\subsection{Continuous Group Key Agreement}
\label{sec:cgka}

A Continuous Group Key Agreement (CGKA) protocol is a scheme that allows a set of users to establish a common secret. This shared value changes dynamically with every modification of the state of the group, whether it be insertions or eliminations of users. In addition to the shared secret, each member possesses some private information such as a signing key pair linked to an identity credential. Each state of the group is called an \textit{epoch}; whenever a member issues a modification to the state of the group through a \textit{commit message}, a new epoch is created with a different shared secret. Throughout this document, we will refer to the group member who generates a commit as \textit{committer}. CGKA protocols are designed to resist the leakage of the shared secret by providing both Forward Secrecy (FS) and Post-Compromise Security (PCS) across epochs.

A commit may add a new member to the group or remove an existing one. Additionally, members can modify their individual state if they believe it has been compromised. This also refreshes the group's state and may modify some of the group's contextual information and parameters. 

Efficiency is also a main focus of CGKA protocols, aiming to achieve logarithmic complexity in updates to the state of the group~\cite{bounds}. To this end, tree structures in which each user possesses information stored in a leaf node~\cite{art,treekem} are common in the literature. 

\subsection{Messaging Layer Security (MLS)}

MLS uses TreeKEM \cite{treekem} as its underlying CGKA protocol. The group's state is structured as a binary tree in which every node holds a cryptographic key pair. The leaf nodes represent the members and contain other information like their credentials and signature key. The participants only know the secret key of those nodes that are included in their path to the root. The secret contained in the root node is known to all members, and thus it is employed to derive the shared secret.

Commits that alter the group's state may modify both the committer's leaf node and the secret key of intermediate nodes in its path to the root. In order to securely transmit these changes to other users, a set of path secrets is generated by encrypting the new secrets with some of the tree's public keys, such that all members can recalculate the shared secret.

For the correct functioning of an MLS group, two abstract services are defined \cite{mls_arch}: the Authentication Service (AS) and the Delivery Service (DS). The former is tasked with generating the member's credentials and assisting in validating the identity of other members. The Delivery Service is tasked with distributing messages and storing relevant information about both groups and users. 

\subsection{Communication in MLS}

The following message types are defined in MLS \cite{mls}:

\begin{itemize}
    \item Handshake: Messages that apply or propose changes to the group's state. Handshake messages can either be \textit{Proposals} that communicate the intention to apply a change to the group but do not directly apply it; or \textit{Commits} that contain one or more proposals and apply them to the group. 
    \item Welcome: Generated by commits that include new members to the group. It contains a copy of the group's Ratchet Tree and any other information that the new user needs in order to participate in the group.
    \item Application: any message exchanged between group members, protected by the group's secret.
\end{itemize} 

These messages are generated by users and distributed to members of the group (in the case of the Welcome message, to the new member) through the Delivery Service. 

In addition to delivering protocol messages, the DS must also store some information generated by users and groups such that it can be accessed by any party on demand. We will refer to this abstract functionality of the DS as \textit{Directory}. The Directory stores \textit{KeyPackages}, which are generated by users and contain cryptographic information required to insert them into an MLS group. They are included in Add proposals. Group Information (\textit{GroupInfo}) packages are also stored in the Directory, and they contain the group's Ratchet tree and other contextual information that is required for performing External Joins. 

\subsection{Group Updates}

\begin{figure*}
    \centering
    \begin{subfigure}[b]{0.45\textwidth}
        \centering
        \includegraphics[width=1\linewidth]{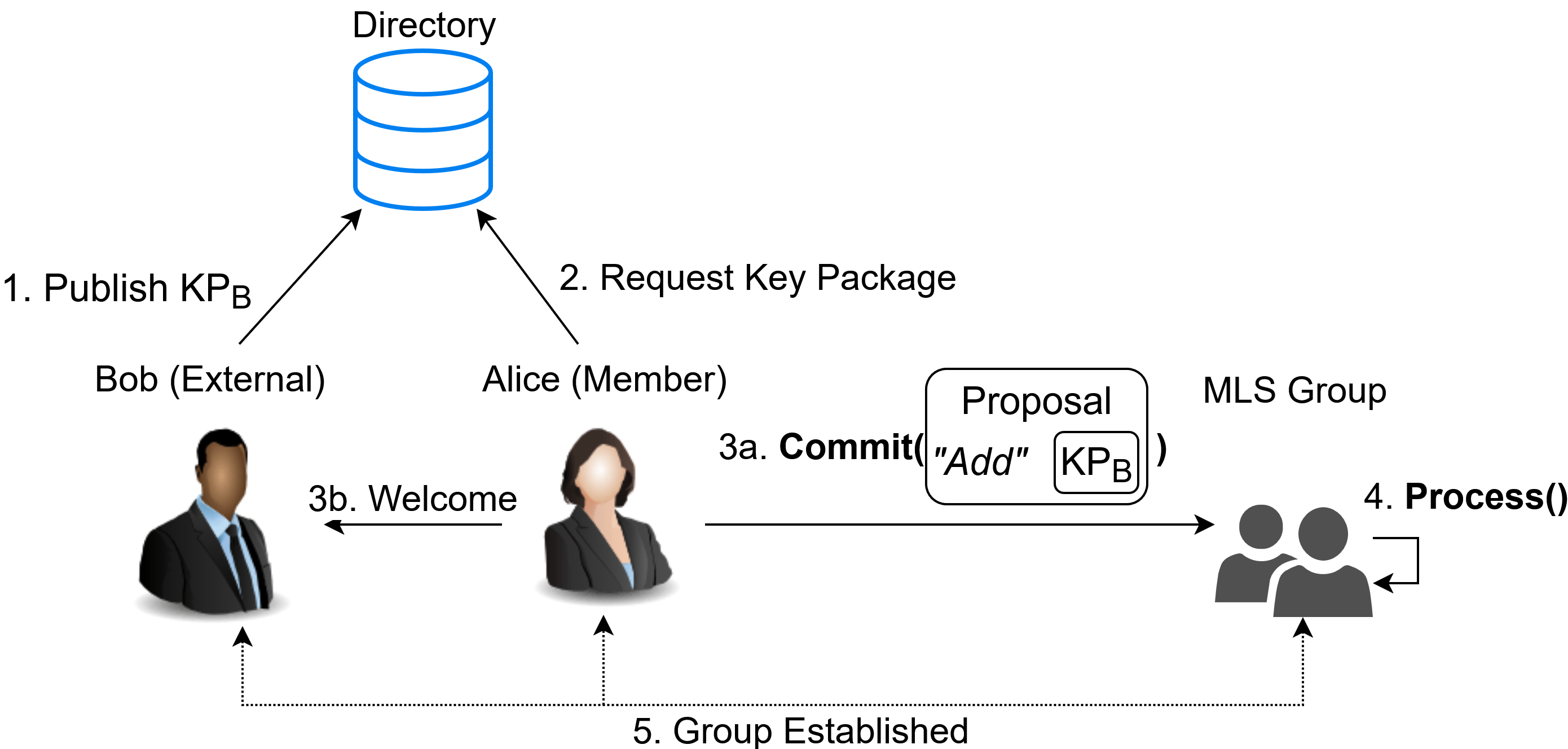}
        \caption{Invitation.}
        \label{fig:new_add}
    \end{subfigure}
    \begin{subfigure}[b]{0.45\textwidth}
        \centering
        \includegraphics[width=1\linewidth]{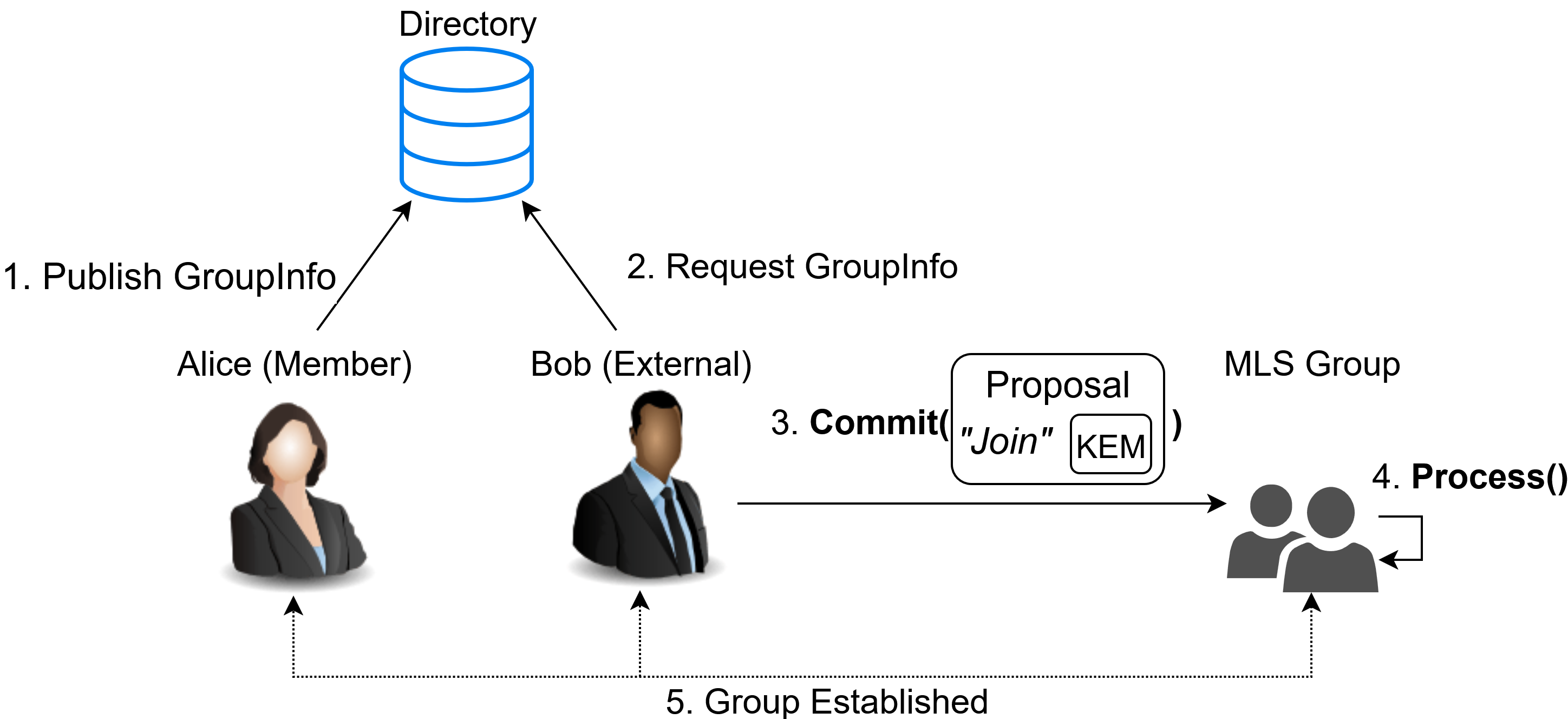}
        \caption{External Join.}
        \label{fig:new_join}
    \end{subfigure}
    \caption{Overview of the Invitation and External Join methods through which new members are added to the MLS group.}
    \label{fig:new}
\end{figure*}

As mentioned, a group's state is modified by commits which can either add or remove members or update their individual state. Although the two latter operations are fairly simple, inserting new members into an MLS group requires additional steps. They can either be added through Invitation ---if a current member inserts them into the group--- or External Join ---if they insert themselves \cite{cgka_analysis_etk}.

Figure \ref{fig:new_add} shows the message flow for inviting new users to the group. As a prerequisite, the new user Bob must have uploaded a Key Package created by him to the Delivery Service's Directory. Then, the group member member Alice obtains it and generates an Add proposal that includes the Key Package. When the proposal is committed, Alice creates a Welcome message directed to Bob, which contains the information needed to participate in the group. All other members of the group then process Alice's commit.

In contrast to Add proposals, External Joins are initiated by the joiner, who is not a member of the group. The procedure is shown in Figure \ref{fig:new_join}: First, the joiner Bob accesses the group's GroupInfo, which was uploaded to the Directory by the group member Alice. Then, he generates a value that will be used to derive the new epoch's shared secret and includes it in a Join proposal which then committed by Bob and processed by all members of the group.

\section{Testbed Design} 
\label{sec:env}

In this Section we describe the scope of our experimental analysis of the performance of MLS. We start by introducing the experiments that study different aspects of the evolution of an MLS group. Then, we discuss the measurements we will extract from our each of the experiments.

\begin{figure}
\centering
\includegraphics[width=0.8\linewidth]{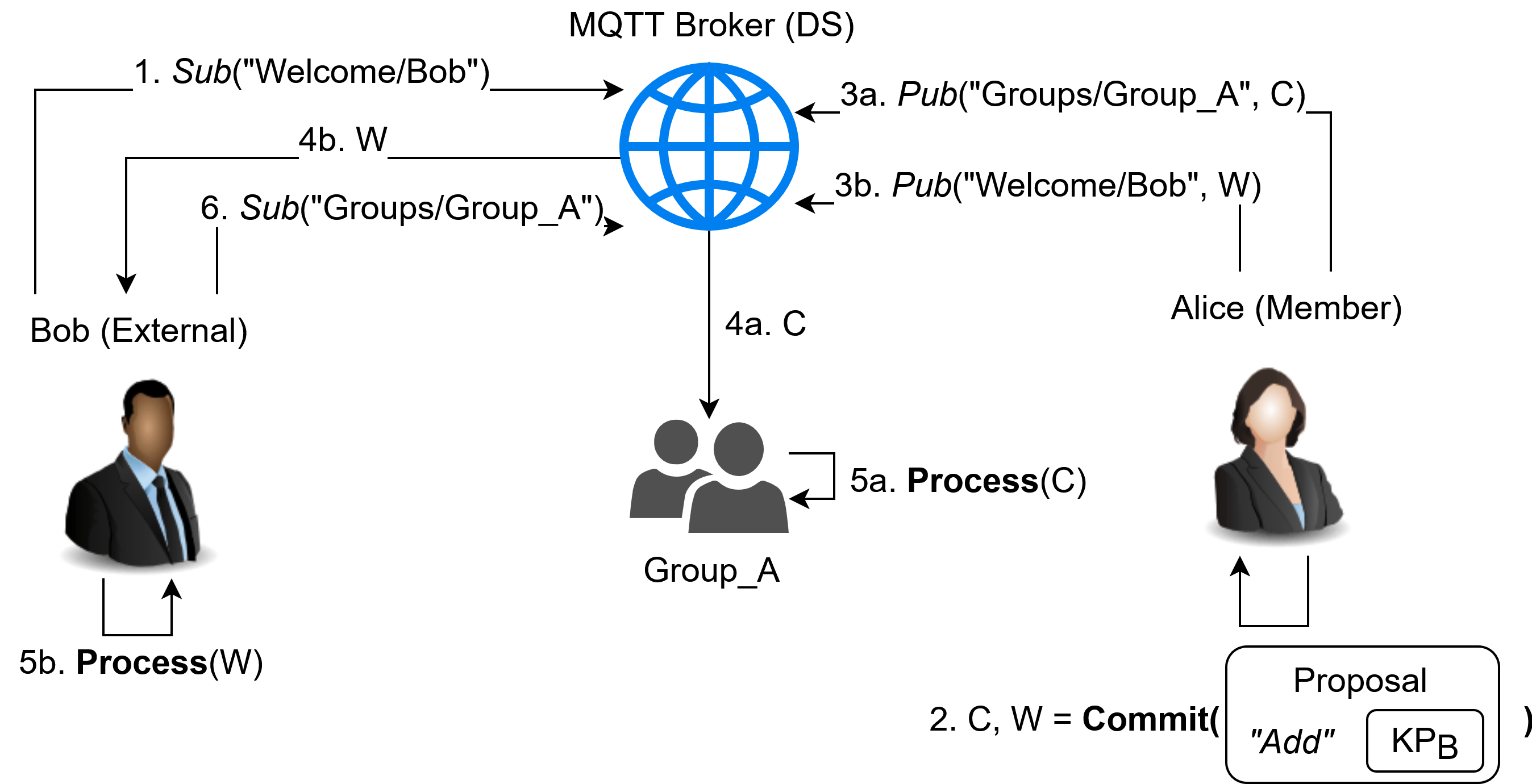}
\caption{Publish/Subscribe architecture in our testbed.}
\label{fig:pubsub}
\end{figure}

Our testbed is composed of two elements. Firstly, a number of identical \textit{emulated MLS clients} that independently participate in MLS groups by creating or joining them and issuing and processing updates. Secondly, a centralised Publish/Subscribe Delivery Service that forwards messages to group members. In this architecture, users can publish messages to different topics such that they are received by all users that are subscribed to said topic. Figure \ref{fig:pubsub} shows the handling of topics in the testbed. Every group has a different topic to which all current members are subscribed. Every user is subscribed to a \textit{welcome} topic through which it receives all Welcome messages directed to them. Additionally, all groups have their own topic that all members are subscribed to, and every handshake and application message of that group is published through it. 

As mentioned, our testbed aims to study the empirical performance of MLS under different situations. To that end, we have defined a series of experiments, that are characterised by the specific configuration that is shared by all emulated clients. In each of them we aim to study the evolution of the group under specific conditions. Table \ref{tab:experiments} shows a summary of the experiments to be performed, which will be detailed in this Section.

\begin{table*}
\centering
\resizebox{\textwidth}{!}{
\begin{tabular}{l|l|l|l|l|l}
\toprule
Experiment & Batch & Paradigm       & Committer & Target users & Measurements                       \\ \midrule

\textit{Commit}                           & \multirow{5}{*}{Group Evolution: Paradigm}           & Commit                             & \multirow{6}{*}{Random} & \multirow{7}{*}{10000}            & \makecell[l]{Gen and Proc Time \\ Commit, Welcome and KP Size \\ Latency} \\ \hhline{|-~-~~-|}
\textit{2 Prop}                           &                                                      & \multirow{3}{*}{Propose-and-Commit} &                         &                                & \multirow{4}{*}{\makecell[l]{Gen and Proc Time \\ Commit Size}}               \\ \hhline{|-~~~~~|}
\textit{4 Prop}                           &                                                      &                                     &                         &                          &                               \\ \hhline{|-~~~~~|}
\textit{8 Prop}                           &                                                      &                                     &                         &                          &                               \\ \hhline{|-~-~~~|}
\textit{External Join}                    &                                                      & External Join                       &                         &                         &                                \\ \hhline{|----~-|}
\textit{First}                            & \multirow{2}{*}{Group Evolution: Tree Degradation} & \multirow{5}{*}{Commit}             & First                   &                         &       \multirow{2}{*}{\makecell[l]{Gen and Proc Time (Deep) \\ Commit Size}} \\ \hhline{|-~~-~~|}
\textit{Last}                             &                                                      &                                     & Last                    &                         &                                                         \\ \hhline{|--~---|}
\textit{100 Users}                        & \multirow{3}{*}{Stable Group Size}                   &                                     & \multirow{3}{*}{Random}  & 100                      &   \multirow{3}{*}{\makecell[l]{Gen and Proc Time \\ Commit Size}}                                                      \\ \hhline{|-~~~-~|}
\textit{750 Users}                        &                                                      &                                     &                         & 750                      &                                                         \\ \hhline{|-~~~-~|}
\textit{5000 Users}                       &                                                      &                                     &                         & 5000                      &                                                         \\ \bottomrule
\end{tabular}}
\caption{Summary of the performed experiments.}
\label{tab:experiments}
\end{table*}

\subsection{Group Evolution}
\label{sec:evo}

Our first batch of experiments explores how measurements are affected by the evolution of the MLS group as it grows in size. All experiments described in this section will consist on the emulation of an MLS group to which new members are constantly being added. The operation to perform and the user that will perform it will depend on the specific experiment. 

\subsubsection{Group Updates Paradigm}

In MLS, the following paradigms can be employed to issue updates to the group, including adding new members:

\begin{itemize}
    \item \textit{Commit}: Modifications are directly and individually applied to the group by a Commit.
    \item \textit{Propose-and-commit}: A single commit includes and applies any number of previously exchanged proposals, possibly generated by different members. This effectively reduces the amount of commits generated while maintaining the overall behaviour of the group.
\end{itemize}

The \textit{propose-and-commit} paradigm reduces the number of modifications to the Ratchet Tree that are required to perform the same functionality. However, it also introduces an additional overhead: the generation, distribution and processing of proposal messages. 

We will also consider External Joins in our discussion of paradigms. Although External Joins are not incompatible with any of the other two paradigms ---as removes and updates can still be either \textit{commit} or \textit{propose-and-commit}--- they do represent a different method for adding new members to the group that can be compared with them.

To study the impact of the chosen paradigm in MLS, we will conduct a total of five experiments that differ on how updates are issued and how new members are inserted into the group: one for the \textit{commit} paradigm, one for External Joins and three for the \textit{propose-and-commit} paradigm using different amounts of proposals per commit: 2, 4 and 8. We will refer to these experiments as \textit{Commit}, \textit{External Join}, \textit{2 Prop}, \textit{4 Prop} and \textit{8 Prop}, respectively.

\subsubsection{Tree Degradation}

\begin{figure}
    \centering
    \begin{subfigure}[t]{0.45\columnwidth}
        \centering
        \includegraphics[width=1\columnwidth]{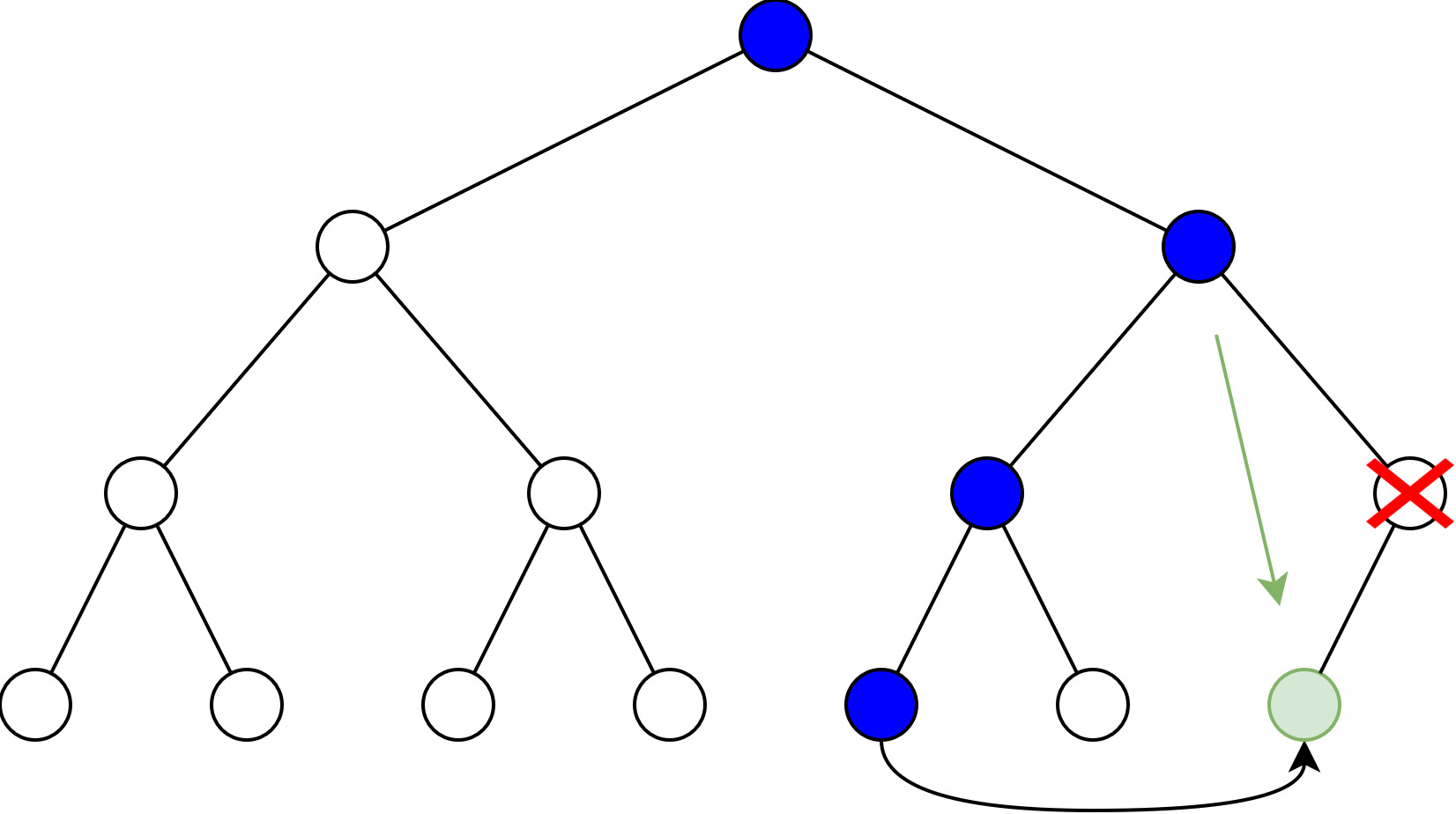}
        \caption{A member adds another user, which creates new empty nodes to the tree. As the committer cannot know the private keys of the new nodes, they remain empty.}
        \label{fig:tree_add}
    \end{subfigure}
    \hfill
    \begin{subfigure}[t]{0.45\columnwidth}
        \centering
        \includegraphics[width=1\columnwidth]{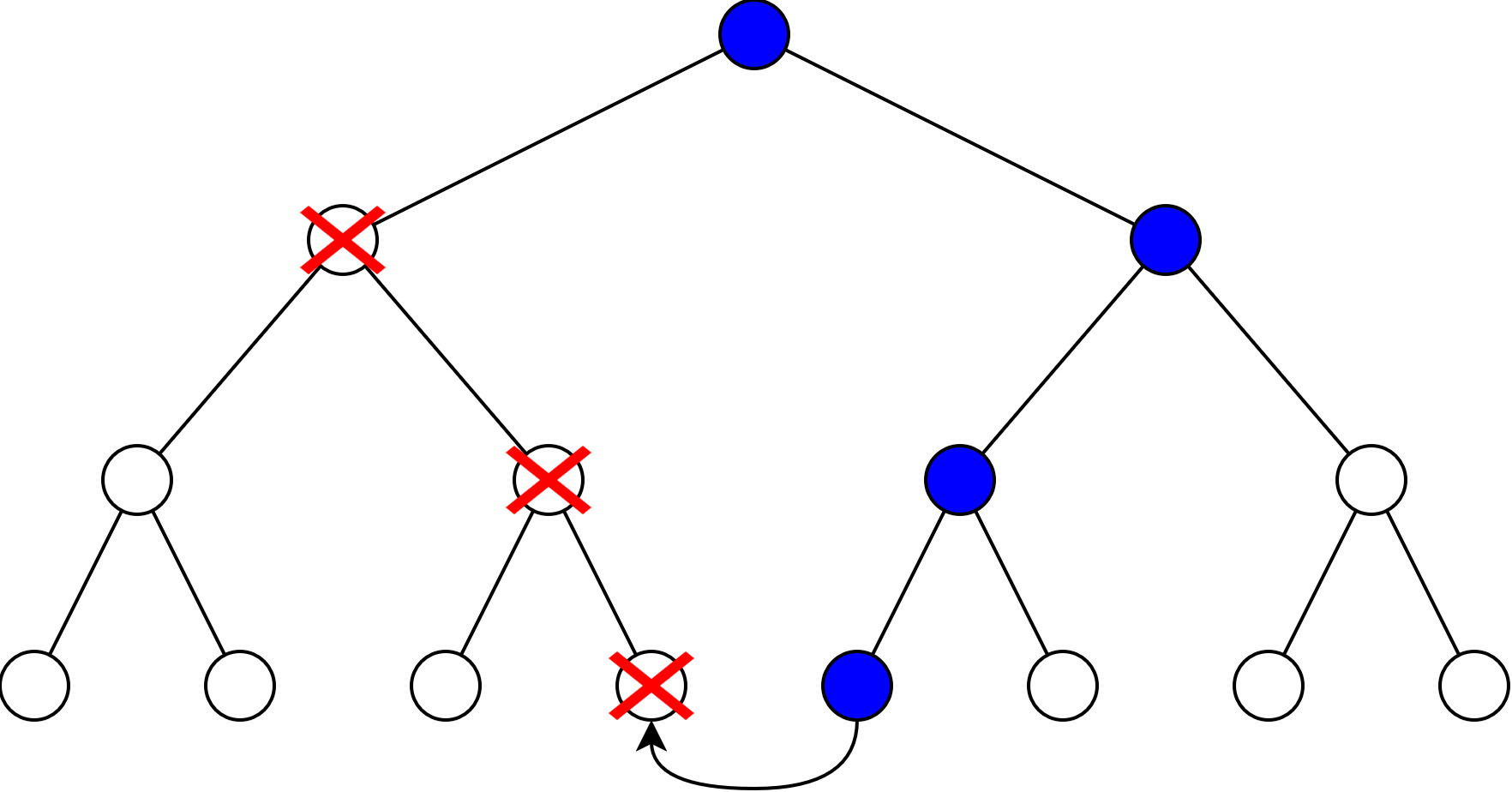}
        \caption{A member removes another user. To ensure the removed user cannot decrypt future updates, all of its known nodes are emptied.}
        \label{fig:tree_remove}
    \end{subfigure}
    \caption{Events that degrade the state of the Ratchet Tree.}
    \label{fig:tree_state_evo}
\end{figure}

\begin{figure}
    \centering
    \begin{subfigure}[t]{0.45\columnwidth}
        \centering
        \includegraphics[width=1\columnwidth]{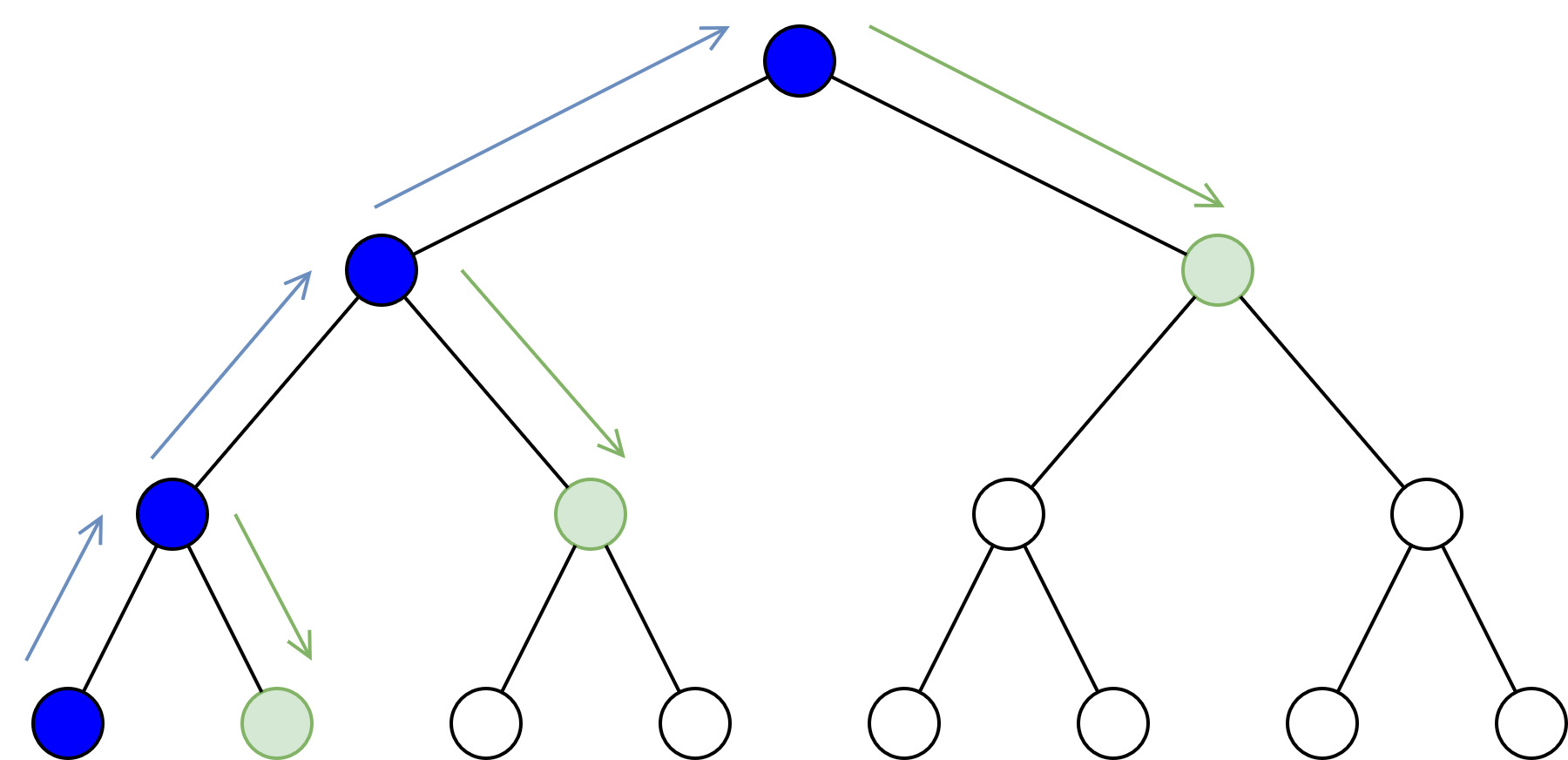}
        \caption{Ratchet tree on its optimal state, with all intermediate nodes populated. New commits includes 3 ciphertexts.}
        \label{fig:tree_opt}
    \end{subfigure}
    \hfill
    \begin{subfigure}[t]{0.45\columnwidth}
        \centering
        \includegraphics[width=1\columnwidth]{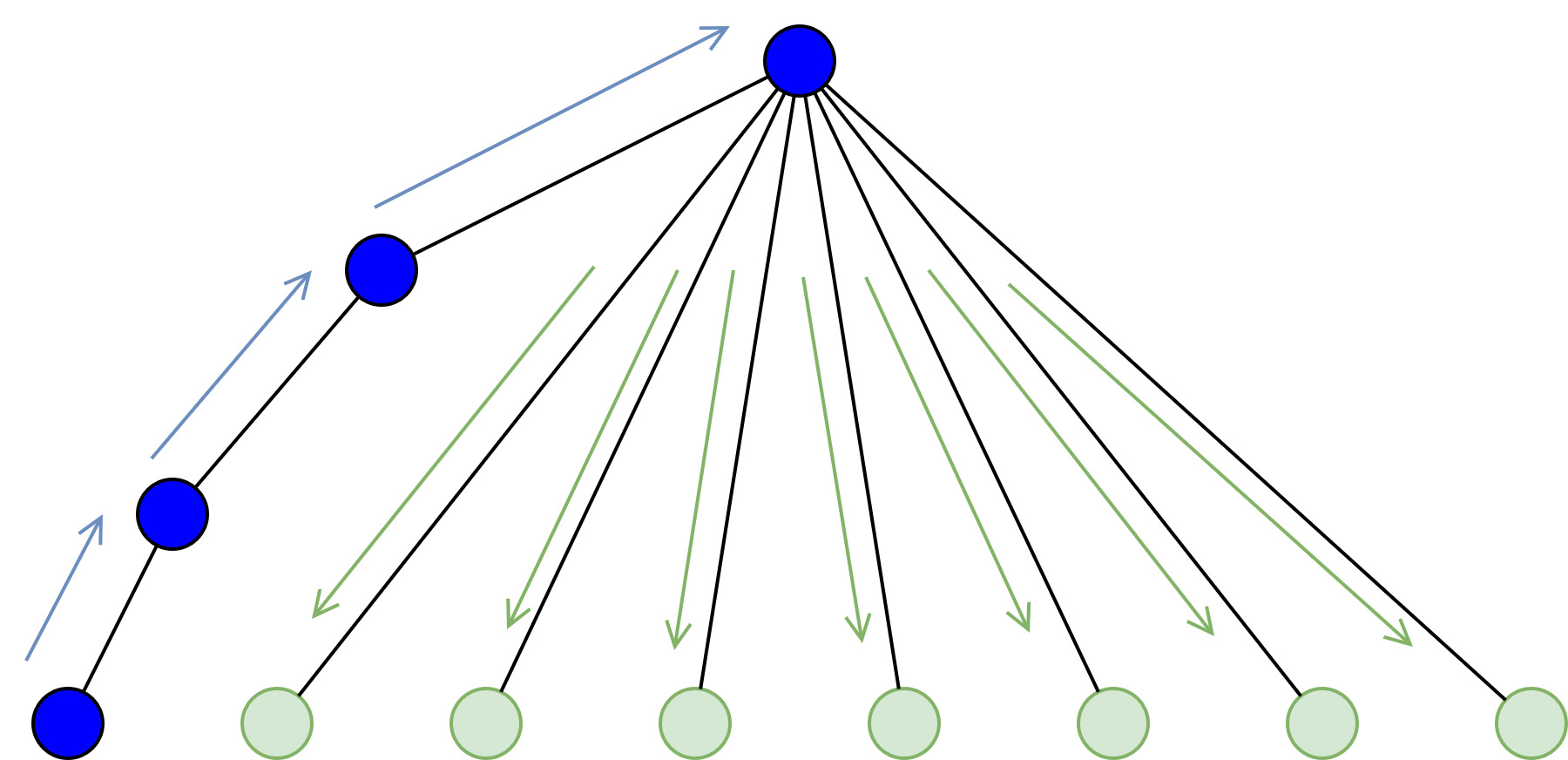}
        \caption{Ratchet tree on its least optimal state, with all intermediate nodes empty. A new commit includes 7 ciphertexts.}
        \label{fig:tree_last}
    \end{subfigure}
    \caption{Impact of tree degradation in the performance of MLS for a 8-member group.}
    \label{fig:tree_state}
\end{figure}

The tree structure of the group state allows MLS to achieve logarithmic efficiency in the best-case scenario. However, this only occurs when all the intermediate nodes are populated. 
Figure~\ref{fig:tree_state_evo} shows two scenarios in which intermediate nodes may become blanked: during additions to the rightmost side of the tree and removals. Blanked intermediate nodes are populated again only when a member whose path to the root includes it issues an update. The impact of blanked intermediate nodes is felt on the amount of ciphertext that need to be distributed, as shown in Figure~\ref{fig:tree_state}. 

We will analyse the importance of the tree degradation by conducting two different experiments. They will measure performance for the worst ---all intermediate nodes empty --- and best --- all intermediate nodes populated--- cases. The worst possible state is achieved if all updates are issued by the first user to join the group, as shown in Figure \ref{fig:tree_add}. Conversely, the best possible state is reached if every user issues an update; we achieve this by ensuring that only the last member to join can issue updates. We refer to this experiments as \textit{First} (worst possible state) and \textit{Last} (best possible state).

\subsection{Stable Group Size}
\label{sec:stable}

Whereas in our first batch of experiments we analyse how the measurements change as the group evolves in size, the second batch focuses on the performance of group operations at static group sizes. Our objective is to measure the variance of the cost of updates at a particular group size. Furthermore, we also analyse if this variance remains stable throughout different group sizes, or if larger groups have more volatile computing costs than smaller ones.

The experiments will unfold as follows: the group evolves normally until it reaches the target group size. Then, members will alternate between removing and adding users, such that the group size remains stable. We will perform three different experiments for target group sizes 100, 750 and 5000 users, as they represent different orders of magnitude close to a power of 2. We refer to this experiments as \textit{100 Users}, \textit{750 Users} and \textit{5000 Users}, respectively.

\subsection{Measurements}

In this work we will focus in computational costs, that is, the time it takes to compute certain MLS operations. In particular, we will obtain generation and processing times for different types of MLS messages, measured in CPU time. Furthermore, we will also analyse the size of MLS message: this includes handshake messages --- both commits and proposals --- as well as other message types like Welcome and GroupInfo packages. 

\paragraph{Update Cost}

When considering the impact of paradigm, we unify all costs --- both time costs and message sizes --- to account for the different amount of proposals being applied in a single commit. Recall that whereas in the \textit{commit} paradigm every handshake message applies exactly one modification, in the \textit{propose-and-commit} paradigm a commit that applies $n$ proposals requires the exchange of $n + 1$ messages: each of the $n$ proposal messages and the commit message itself. Thus, we define the \textit{update cost} as

\[ \mathsf{UC} = \frac{cc + \sum_{i=0}^{n} cp_i}{n}, \]
where $cc$ and $cp_i$ are the costs of the commit and its $i$th proposal, respectively. Note that in the \textit{commit} paradigm, $\sum_{i=0}^{n} cp_i = 0$ and $n=1$, thus the update cost equals $cc$.

\paragraph{Deep Analysis}

The experiments \textit{First} and \textit{Last} respectively represent the least and most optimal states of the Ratchet Tree. As shown in Figure \ref{fig:tree_state}, this impacts the amount of ciphertexts that must be created to communicate the new state of the group. In order to further understand the differences between both experiments, we will additionally perform a deeper analysis for these experiments that subdivides the measurements obtained into categories, each representing a different aspect of the creation of a commit:

\begin{itemize}
    \item \textit{Validation}: Operations to ensure that the proposals to apply are valid. Of particular interest is the MLS requirement that encryption keys be unique among all members.
    \item \textit{Cryptography}: TreeKEM operations such as updating and encrypting the intermediate nodes' secrets and updating the key schedule plus signing and encrypting MLS messages.
    \item \textit{Tree}: Operations related to parsing the Ratchet Tree such as calculating the path to the root and finding which keys to use to encrypt and decrypt path secrets. Additionally, operations that maintain the integrity of the tree, such as and updating the tree hashes and parent hashes every epoch \cite{mls}.
    \item \textit{Welcome}: Operations to communicate the state of the group to a new member, such as exporting the GroupInfo package and the Ratchet Tree. Only applicable to Commit generation.
    \item \textit{Storage}: Operations to provide persistence to the application by saving the group's state into memory.
\end{itemize}

\paragraph{Latency}

Finally, we also measure \textit{latency}, defined as the time between the instant a message is sent by a committer and the instant it is received by a processer. Whenever a commit is issued but has not been processed yet by all members, the group is in an inconsistent state as members disagree on the current epoch and group state: latency determines how long this inconsistency may last. To avoid redundancy, latency does not include neither commit generation nor processing times. 

\section{Emulation Environment}
\label{sec:impl}

In this Section we introduce the implementation of the presented testbed. Our 
code\footnote{Available at \url{https://github.com/SDABIS/mls_experimental_analysis}} consists mostly of Rust code that implements the emulated MLS client, as well as the Docker infrastructure required to execute the Delivery Service and multiple clients. 

\subsection{Delivery Service}

MQTT is a protocol for establishing publish/subscribe queues of messages \cite{mqtt}. The transmission of these messages to all subscribers is managed by a MQTT broker. For our execution environment, we employ the open-source Mosquitto \cite{mosquitto} implementation of a MQTT broker. 

Besides the MQTT broker, the Delivery Service also employs a web server that acts as Directory, with endpoints to store and consume KeyPackages and GroupInfo messages. The latter needs to be updated for each new group epoch, so users aiming to perform an External Join can successfully access the group. Furthermore, this web server also acts as a signaling server in which users can register to notify other participants of their existence, such that they can be invited to MLS groups.

\subsection{Client Emulation}

Our testbed is composed of multiple emulated clients that independently participate in MLS groups. They emulate the behaviour of real clients by randomly adding or removing members, updating their individual leaf nodes and sending application messages. Clients interact with each other through real MLS messages sent through the Delivery Service. 

We have developed our implementation of an emulated client in Rust, using the OpenMLS project \cite{openmls} as a baseline and incrementally expanding upon it. Clients can be deployed in Docker containers, each executing a configurable amount of copies. The behaviour of Clients can be extensively configured to reach specific group states. Each client has a different name and a list of groups they aim to join.

There are two behaviours clients can follow to decide when to act in a group. In the independent mode, clients randomly decide whether or not to send a message at fixed time intervals. This may result in redundant messages if multiple users attempt to commit at the same time. In orchestrated mode, the transcript hash of the group is used as a seed to a RNG, such that all group members agree on a randomly chosen client who issues the next update. This allows for better coordination resulting in faster executions. In either mode, clients always process incoming messages immediately after their arrival. Clients can also be configured to only consider issuing updates if they are the first or last member to join the group, as required by the Tree Degradation experiments described in Section \ref{sec:env}. 

When they are tasked with issuing updates, clients will follow one of the paradigms described in Section \ref{sec:env}. If the \textit{propose-and-commit} paradigm is employed, clients will issue proposals until a threshold of proposals is reached and then commit them. Clients may also perform an External Join to a group in which they are not members.

Clients randomly choose between adds, updates and removals according to configured proportions. In order to reach groups of stable size for the experiments introduced in Section \ref{sec:env}, this proportions can be overridden once the group reaches a threshold of members such that only removes may be issued. Additionally, clients may send randomly-generated application messages. The emulated clients record every action they perform in detailed log file that includes the timestamp and duration of the action.

\section{Experimental Results} 
\label{sec:result}

In this section we deploy the environment presented in previous sections and perform empirical evaluations on its performance. All experiments in Table\ref{tab:experiments} have been executed by providing their respective configuration to the emulated clients. The target group size for all experiments ---except those related to stable group size--- is 10000. We note that in all experiments clients employ the orchestrated behaviour since it allows for faster and more controlled executions. 

All experiments were executed on an Ubuntu 20.04 virtual machine equipped with 128 GB of RAM and 16 CPU cores operating at 3.2 GHz. The 10,000 emulated clients were divided into 10 Docker containers each running 1000 instances. The Delivery Service ---including both the HTTP Directory and the MQTT Broker--- were run in a similar virtual machine located in the same local network.

\subsection{Group Evolution}

In order to ensure groups grew in size, clients were configured with probabilities of 60\% invite, 20\% update and 20\% remove. Every experiment has been executed 3 times; the data shown in this Section are the aggregation of the results obtained in every execution. For each of the following graphs, the 10,000 measurement points have been subsequently aggregated into 100 points, with each point representing the aggregation of 100 measurements. 

\begin{figure}
\centering
\includegraphics[width=0.95\linewidth]{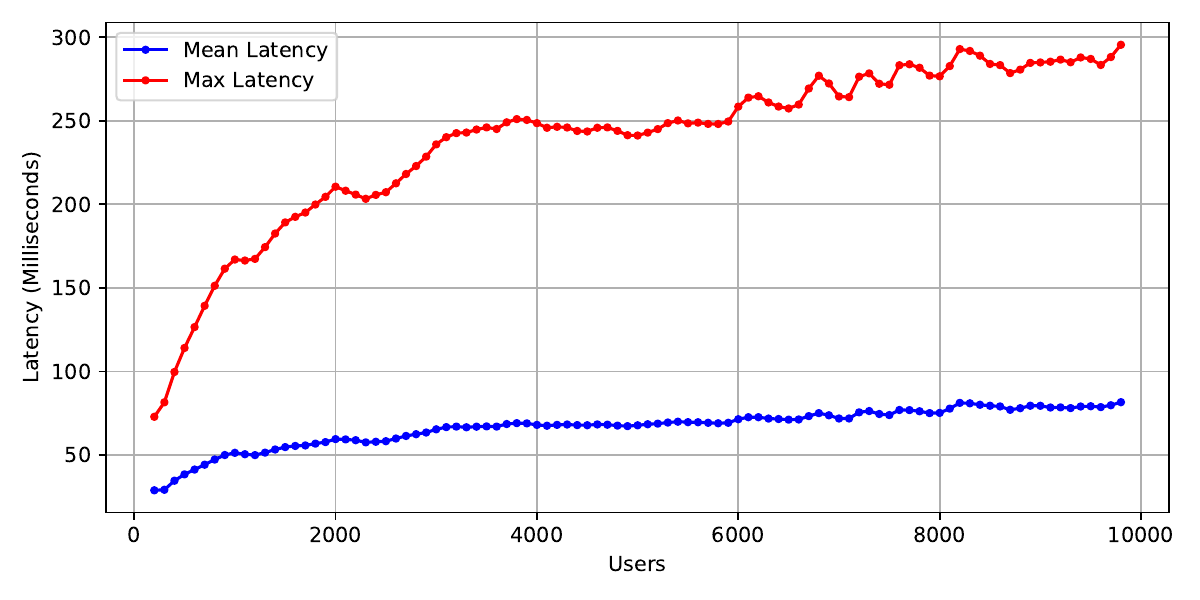}
\caption{Mean and Maximum Latency for the Delivery Service as the number of group members grows.}
\label{fig:lat}
\end{figure}

Figure \ref{fig:lat} shows the mean and maximum latency in the experiment \textit{Commit}. Both measurements increase slowly as the group grows and the consumed bandwidth increases. 

\subsubsection{Paradigm}
\label{sec:paradigm}

We now compare the update cost in our Paradigm experiments. We note that our measurements do not include the cost of retrieving information from the Directory such as KeyPackages for Add proposals, as it remains constant regardless of the chosen paradigm. 

\begin{figure*}[t]
    \centering
    \begin{subfigure}[t]{0.47\textwidth}
        \includegraphics[width=\linewidth]{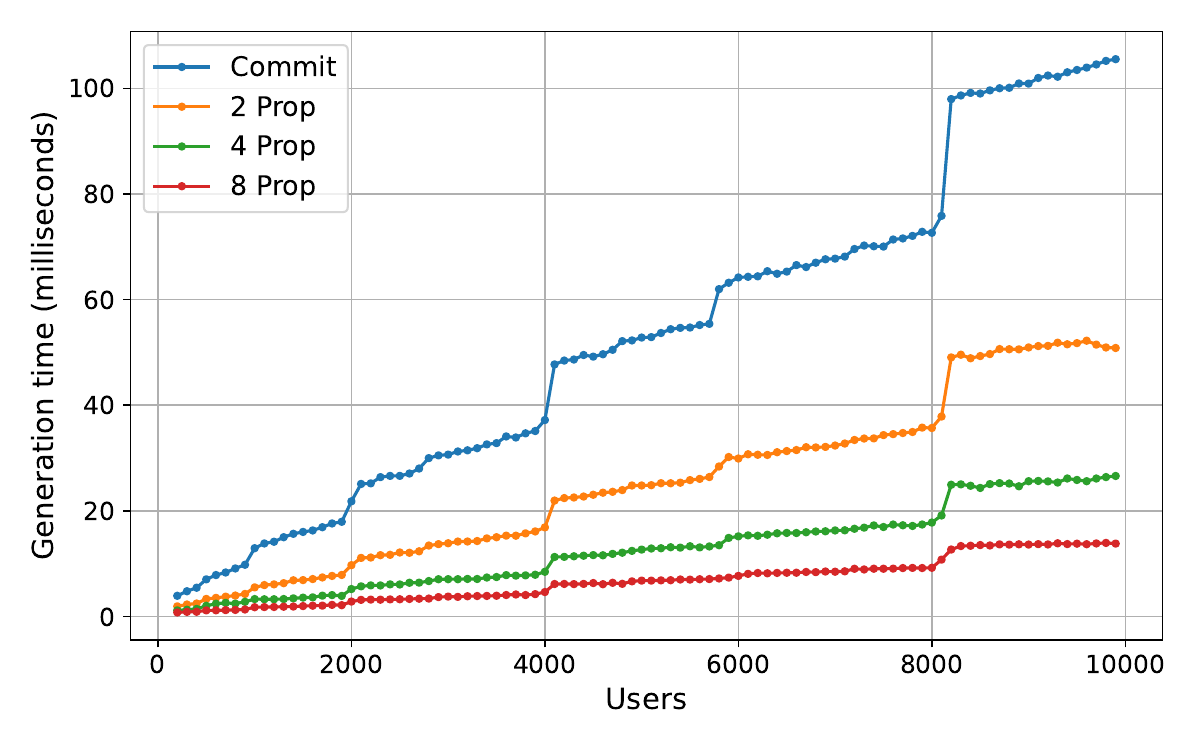}
        \caption{Experiments \textit{Commit}, \textit{2 Prop}, \textit{4 Prop} and \textit{8 Prop}.}
        \label{fig:cpu_gen_para}
    \end{subfigure}
    \hfill
    \begin{subfigure}[t]{0.47\textwidth}
        \includegraphics[width=\linewidth]{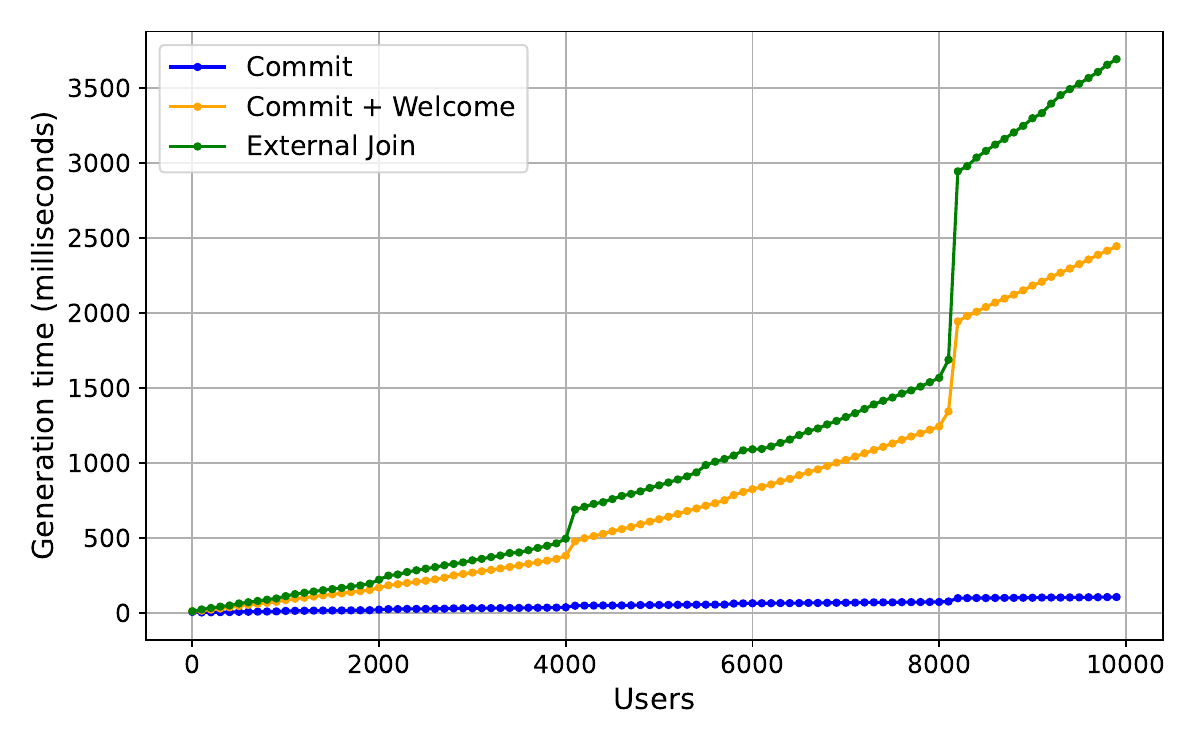}
        \caption{Experiment \textit{External Join}, with \textit{Commit} shown for comparison. Welcome  also including Welcome processing time --- and External Joins.}
        \label{fig:cpu_gen_join}
    \end{subfigure}
    \caption{Update generation time for different paradigms as the number of group members grow. Update cost is calculated by dividing total cost by number of proposals applied in a single commit.}
    \label{fig:cpu_gen}
\end{figure*}

Figures \ref{fig:cpu_gen_para} shows the generation time for the experiments \textit{Commit}, \textit{2 Prop}, \textit{4 Prop} and \textit{8 Prop}, whereas Figure \ref{fig:cpu_gen_join} shows the same measure for the experiment \textit{External Join}. Clearly, the update cost is significantly reduced as the number of proposals increases.
We highlight that an user performing an External Join must also parse the current state of the group including the full Ratchet Tree, which would explain the poor scaling of External Joins. This operation is also performed in an Invitation setting by the invited user when the Welcome message is processed; when accounting for the processing time of Welcome messages, the two methods have a similar cost (as shown in Figure \ref{fig:cpu_gen_join} by the line labelled \textit{Commit + Welcome}).

\begin{figure*}[t]
    \centering
    \begin{subfigure}[t]{0.47\textwidth}
        \includegraphics[width=\linewidth]{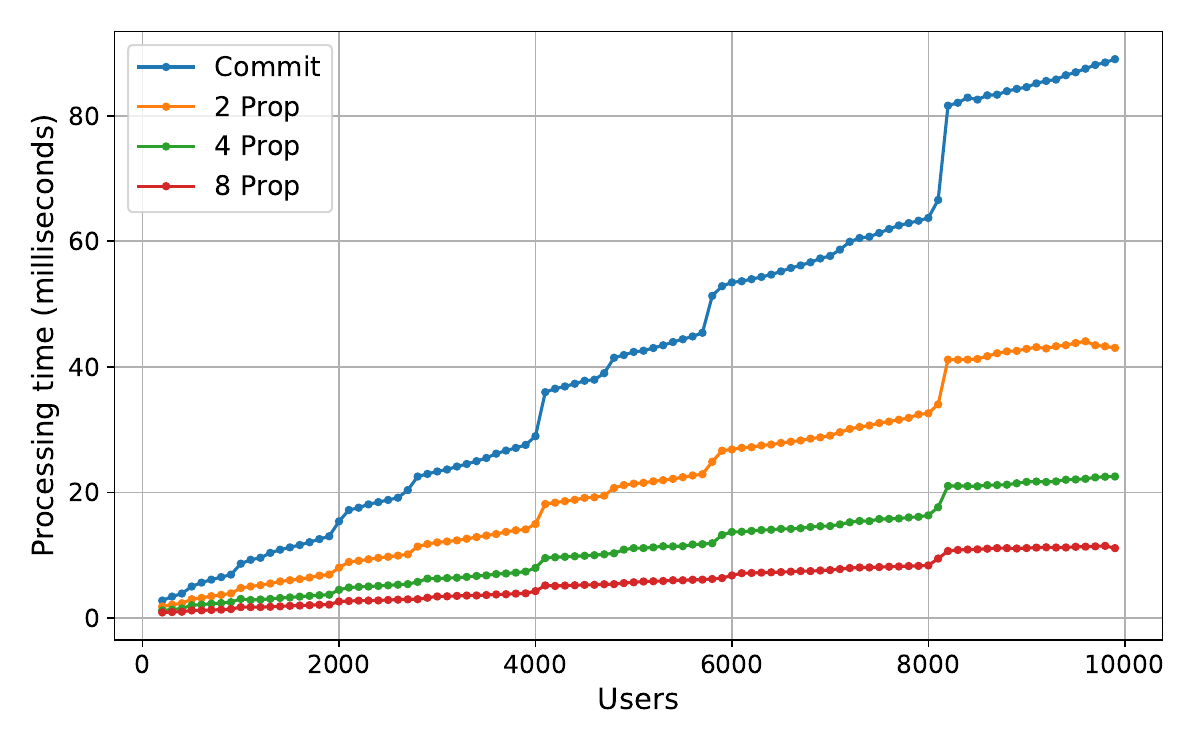}
        \caption{Processing time.}
        \label{fig:cpu_proc}
    \end{subfigure}
    \hfill
    \begin{subfigure}[t]{0.47\textwidth}
        \includegraphics[width=\linewidth]{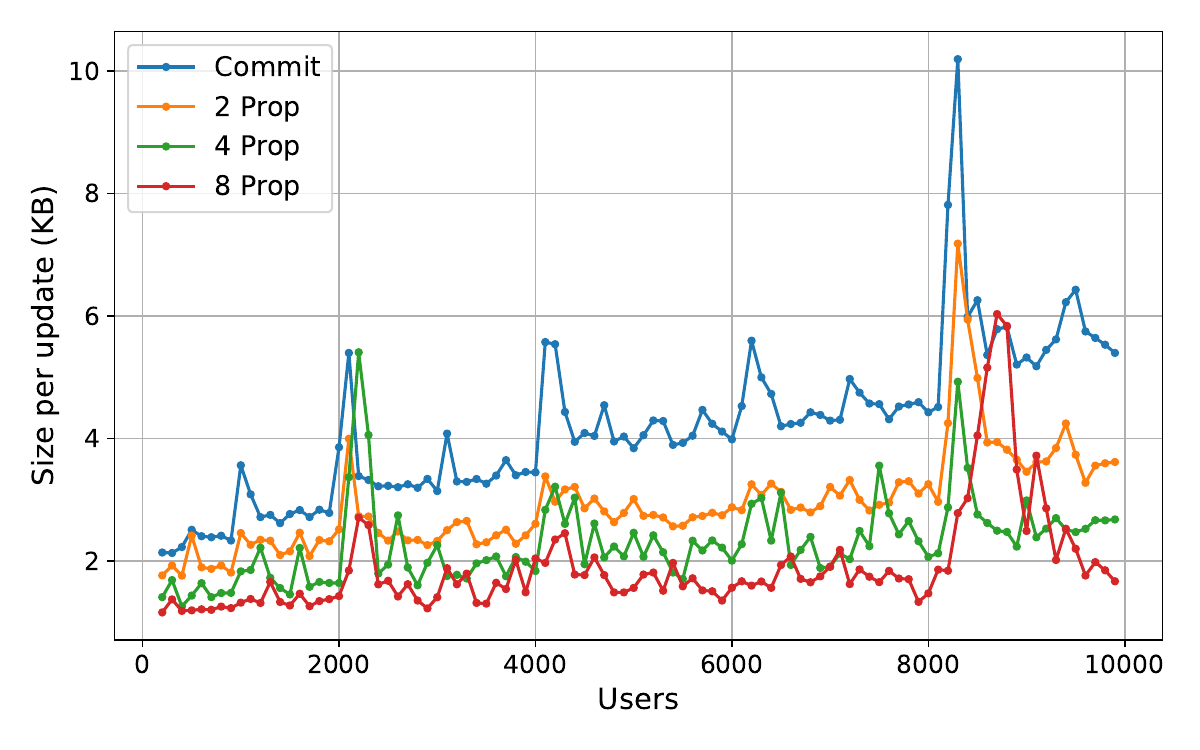}
        \caption{Message size.}
        \label{fig:cpu_size}
    \end{subfigure}
    \caption{Update cost for different paradigms as the number of users grow. Update cost is calculated by dividing total cost by number of updates of proposals applied in a single commit.}
    \label{fig:cpu_cost}
\end{figure*}

Figure \ref{fig:cpu_proc} shows the processing time per Update as the number of users grow. As with generation time, the processing cost is significantly reduced as the number of proposals increases.
Similarly, Figure \ref{fig:cpu_size} shows the message size per update. Clearly, the correlation between the message size and the number of users is much smaller than in the other metrics. The increases at powers of 2 is also noticeable in message size, as shown by the spikes at 4096 and 8192 members. 

\begin{figure}
\centering
\includegraphics[width=0.95\linewidth]{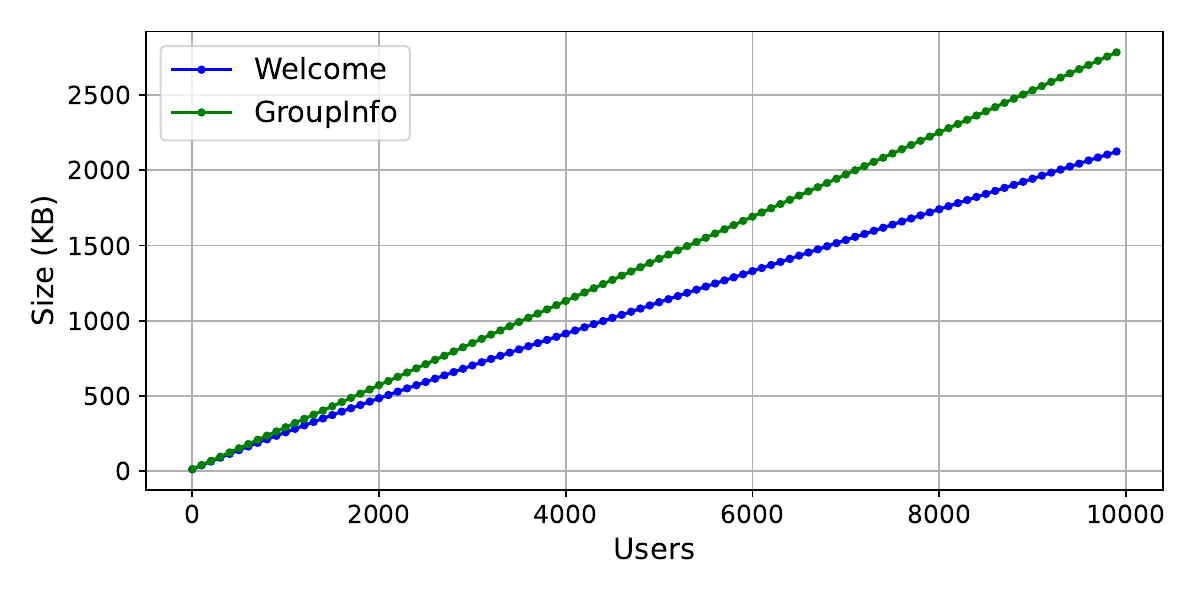}
\caption{Message size in Bytes for Welcome messages and GroupInfo packages.}
\label{fig:size_wel}
\end{figure}

We also evaluate the size of the Welcome messages and GroupInfo packages, which take part in Invitations and External Joins, respectively. Figure \ref{fig:size_wel} shows the obtained results. Since both contain a full copy of the Ratchet Tree, they are significantly larger than any handshake message. 

\subsubsection{Tree Degradation}

\begin{figure}
\centering
\includegraphics[width=0.95\linewidth]{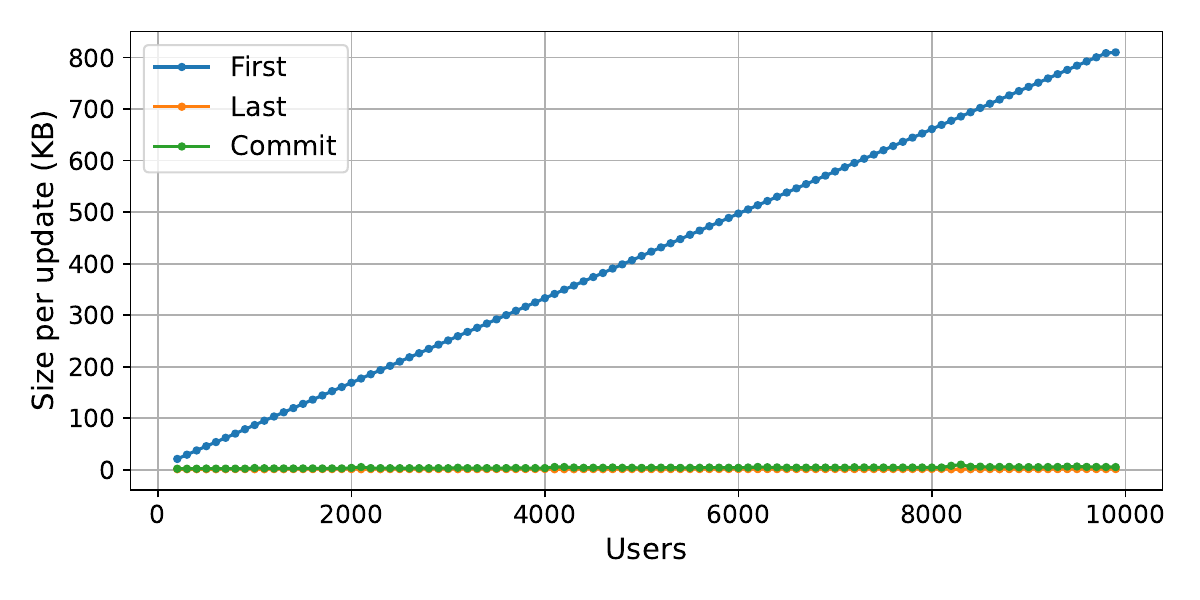}
\caption{Message size for the \textit{First} and \textit{Last} experiments as the number of users grow. The results of the \textit{Commit} experiment are included for comparison.}
\label{fig:state_size}
\end{figure}

Figure \ref{fig:state_size} shows the message size for the experiments \textit{First} and \textit{Last}. Since in the former almost all intermediate nodes are blanked, the commit messages must contain one ciphertext for every member and thus their size grows linearly.

\begin{figure*}[t]
    \centering
    \begin{subfigure}[t]{0.47\textwidth}
        \centering
        \includegraphics[width=\linewidth]{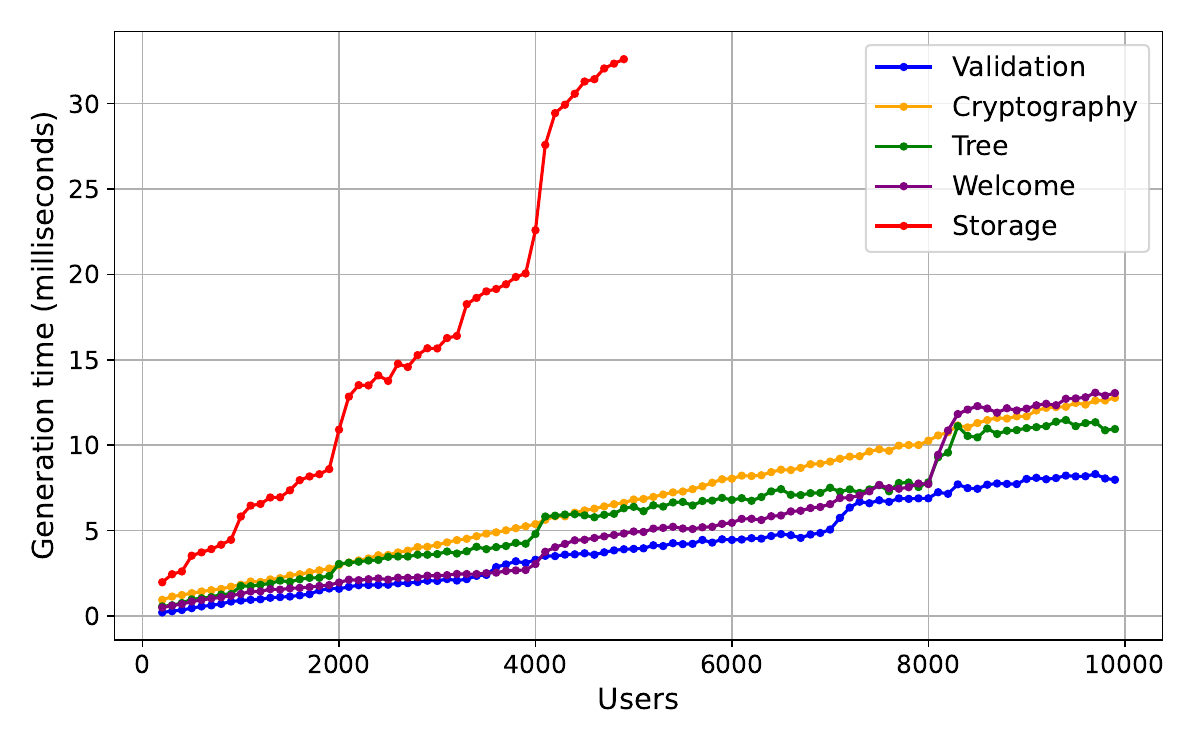}
        \caption{Experiment \textit{First}.}
        \label{fig:state_gen_first}
    \end{subfigure}
    \begin{subfigure}[t]{0.47\textwidth}
        \centering
        \includegraphics[width=\linewidth]{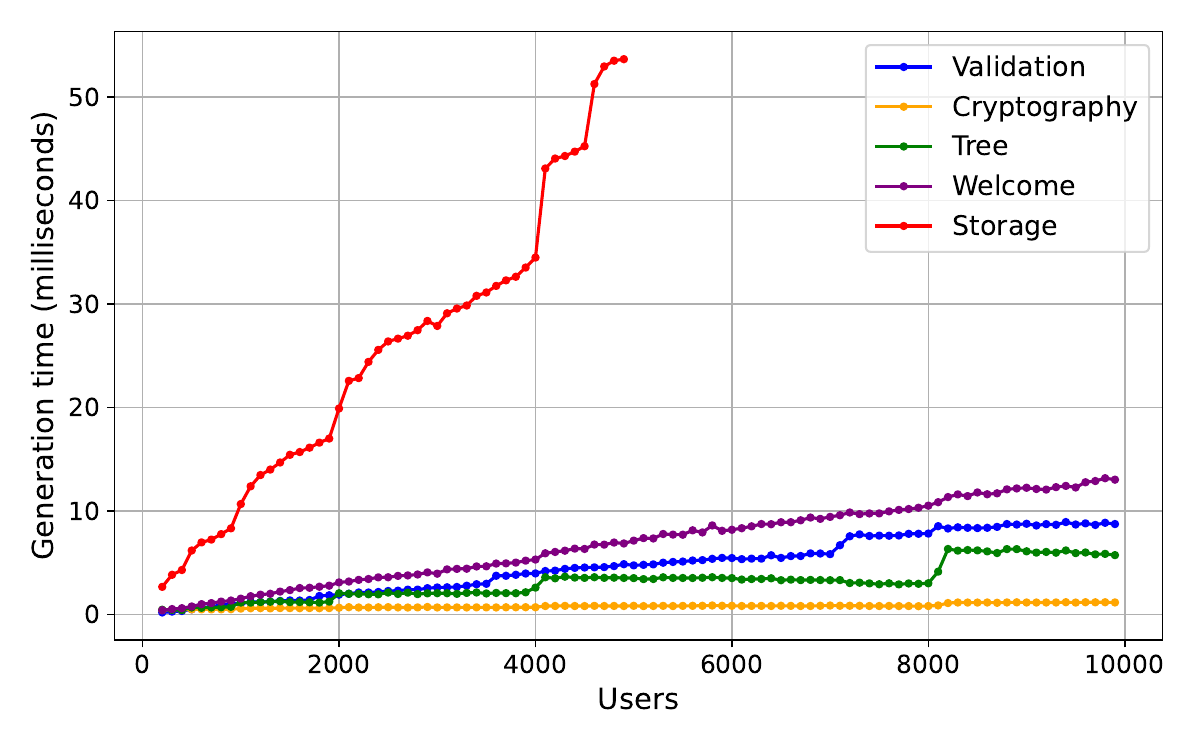}
        \caption{Experiment \textit{Last}.}
        \label{fig:state_gen_last}
    \end{subfigure}
    \caption{Generation time measured for the experiments \textit{First} and \textit{Last} that respectively correspond to the worst and best possible Ratchet Tree states. Measures are subdivided into categories. \textit{Storage} is cropped to improve visibility, as it increases linearly and experiences significant jumps at powers of two.}
    \label{fig:state}
\end{figure*}


Figure \ref{fig:state} shows the distribution of time cost across different categories for commit generation in both experiments. Whereas most categories predictably exhibit the same behaviour, \textit{Cryptography} is significantly more costly in \textit{First} than in \textit{Last}, as the path secrets need to be encrypted once for every member. In both, Processing time is omitted as the cost distribution across categories was identical to that of commit generation. Notably, \textit{Cryptography} represents a small percentage of the total cost: at 10,000 users, it reaches 1\% for the experiment \textit{Last} and 10\% for \textit{First}.

\subsection{Stable group size}

\begin{figure*}[t]
    \centering
    \begin{subfigure}[t]{0.3\textwidth}
        \centering
        \includegraphics[width=\linewidth]{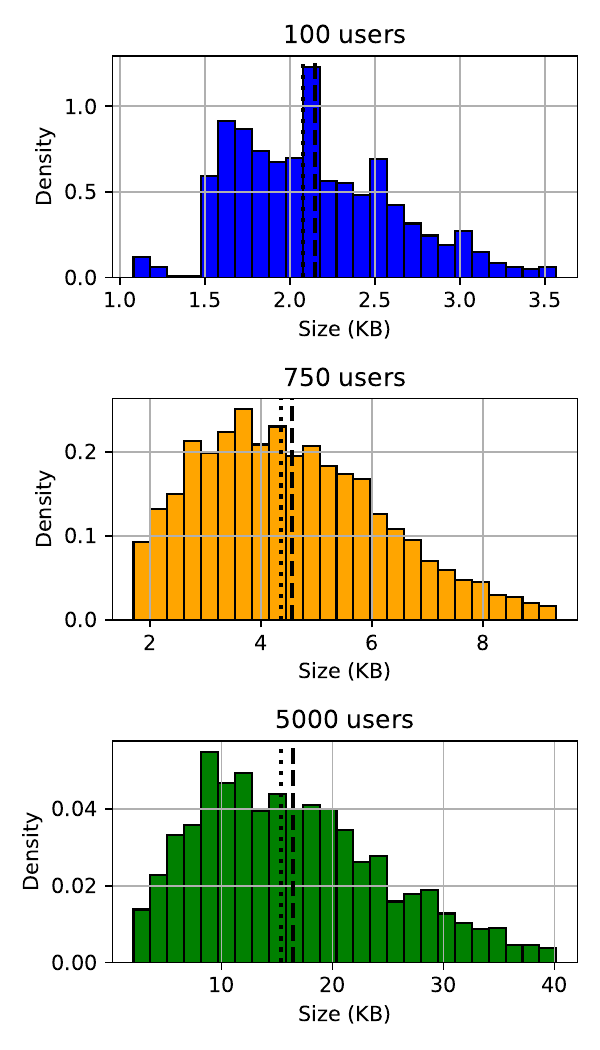}
        \caption{Commit size.}
        \label{fig:stable_size}
    \end{subfigure}
    \hfill
    \begin{subfigure}[t]{0.3\textwidth}
        \centering
        \includegraphics[width=\linewidth]{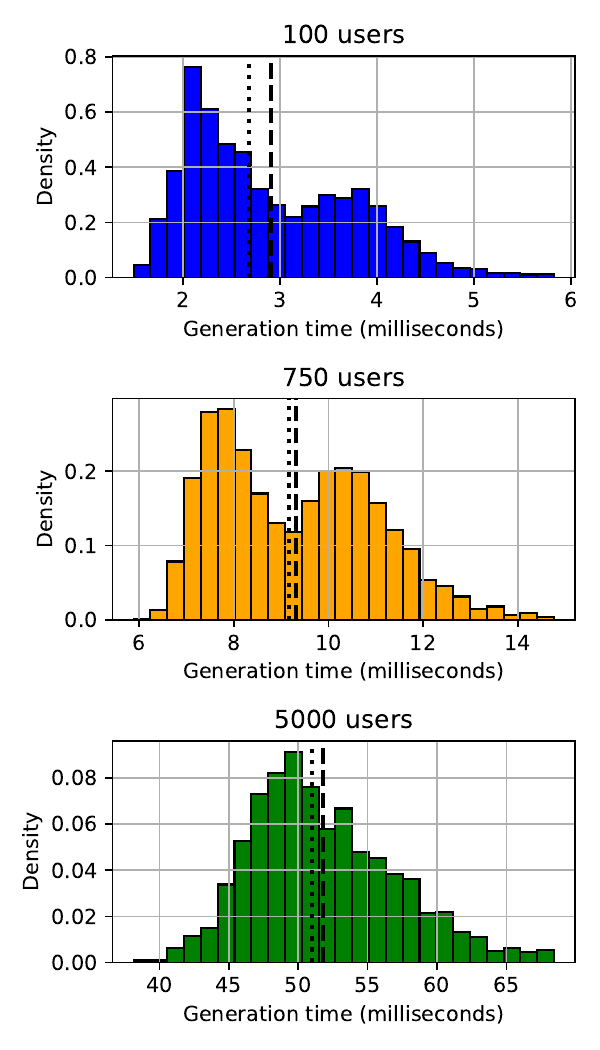}
        \caption{Generation time.}
        \label{fig:stable_gen}
    \end{subfigure}
    \hfill
    \begin{subfigure}[t]{0.3\textwidth}
        \centering
        \includegraphics[width=\linewidth]{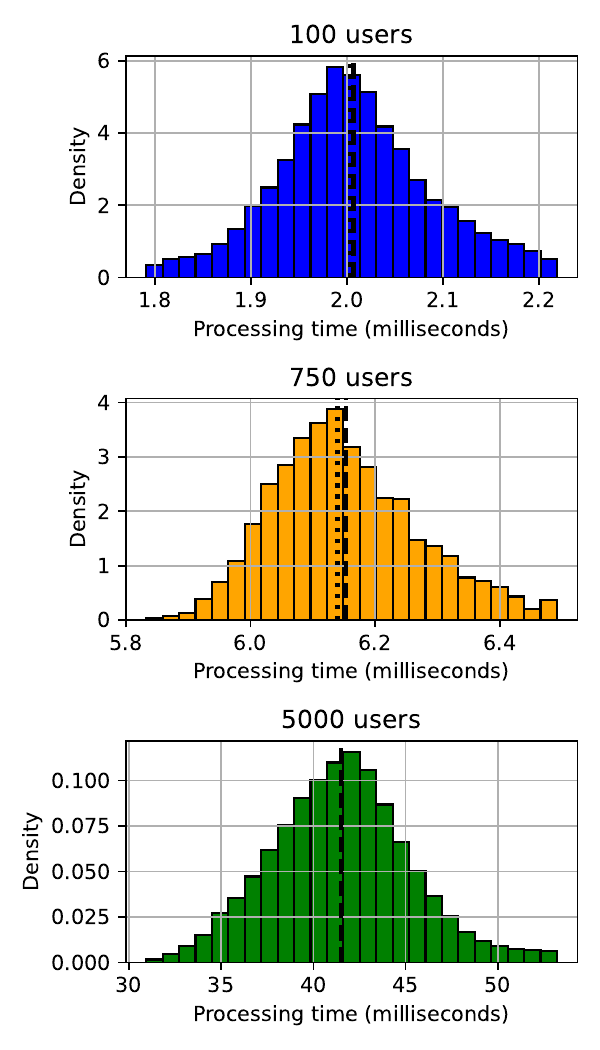}
        \caption{Processing time.}
        \label{fig:stable_proc}
    \end{subfigure}
    \caption{Commit size and generation and processing time for groups of 100 (top), 750 (middle) and 5000 (bottom) members. The dotted and dashed lines respectively represent the median and mean of each series.}
    \label{fig:stable}
\end{figure*}

In this Section we execute the experiments related to stable group size: \textit{100 Users}, \textit{750 Users} and \textit{5000 Users}. For each experiment we measure 10,000 commits at the corresponding group size. Figure \ref{fig:stable} shows the measured commit generation and processing time and message size. The latter metric experiences a noticeably higher variance than the others. In particular, commit processing is the more stable metric. All of them follow a normal distribution, which is more easily observable for larger groups.

\section{Discussion}
\label{sec:discussion}

In this Section we analyse the data introduced in the previous Section and discuss its real-world implications. We also compare our work to other analysis of performance of MLS or other CGKA variants. Finally, we discuss the limitations encountered and propose future work.

\subsection{Analysis}

\textbf{Message cost.} Generation and processing times are comparable and represent a small cost in the order of milliseconds. The \textit{propose-and-commit} paradigm also significantly affects performance: the cost of generating and processing proposal messages is negligible compared to the cost of applying an update to the group. As shown, more proposals per update increase the efficiency of the protocol. The size of handshake messages is less affected by the amount of group members and the paradigm, as it is mostly determined by the current state of the Ratchet Tree. External Joins combine the operations of generating a commit and processing a Welcome message, as the joiner needs to parse the full Ratchet Tree in order to join. When taking these two operations into account, the cost of an Invitation is similar to that of an External Join. 

As shown in Figure \ref{fig:stable}, commit messages at a given group size can have significantly different sizes. This is determined by the number of ciphertexts contained in them, which in turn depends on the number of blanked intermediate nodes as explained in Figure \ref{fig:tree_state}. Notably, commit generation and processing times have less variance as their cost is mostly determined by other operations unrelated to the cryptographic component of MLS.

The increased bandwidth of larger groups has a noticeable impact on latency, as the cost of distributing messages between all members increases. However, this metric grows slower than commit generation and processing times. When comparing Figure \ref{fig:lat} with Figures \ref{fig:cpu_gen} and \ref{fig:cpu_cost}, mean latency is around 2.5 times more expensive than commit generation and processing at 1000 members, whereas their cost is comparable at 10,000 members. 

\textbf{Ratchet Tree Size.} Our results show that the overwhelming majority of the costs both in generation and processing are caused by the size of the full Ratchet Tree. Indeed, Figure \ref{fig:state}'s categories \textit{Storage}, \textit{Welcome}, \textit{Validation} and \textit{Tree} all scale linearly with the size of the group's state; only \textit{Cryptography} is independent, but its cost is negligible in comparison. The result is that the overall scaling of MLS's performance is linear, which contradicts the theoretical results which point to logarithmic performance \cite{cost_1, cost_2, bounds}. We remark that \textit{Storage} cannot be considered optional in practical scenarios, as it is required to provide persistence between sessions. Even then, the remaining categories still scale linearly and their cost cannot be circumvented: \textit{Tree} includes the calculation of tree hashes and parent hashes that prevent insider attacks \cite{itk} and \textit{Validation}'s check that all keys should be different is needed to ensure Post-Compromise Security.

Furthermore, the most expensive operations of the protocol are processing a Welcome and performing an External Commit, as shown in Figure \ref{fig:cpu_gen_join}. Both require processing the full Ratchet Tree, whether it be from a Welcome message or a GroupInfo package. Thus, the most relevant factor to consider regarding performance in MLS is the size of the Ratchet Tree, which scales linearly by the number of members. The logarithmic complexity of the cryptographic operations only applies to a negligible step of the computation.

\subsection{Real-world Implications}

Figure \ref{fig:lat} shows that maximum latency can reach up to 0.3 seconds, which could potentially be much higher depending on the network or Delivery Service employed. This measure determines the amount of time in which the group is in an inconsistent state, since some members have not yet joined the current epoch. Whereas for most applications this span is too short to be noticed, it is something that must be taken into account for situations that require strict synchronisation among group members.

The results of the experiment \textit{First} show that groups in which only a small number of users generate updates are less efficient, with a particularly high commit size. This is significant as real-world messaging applications usually limit updates to one or more administrators. Thus, members should be allowed to periodically update their own leaf node to prevent the accumulation of blanked intermediate nodes.

As shown in Figure \ref{fig:state}, the size of the Ratchet Tree has a significant impact in performance as this structure needs to be managed to maintain its integrity and stored in memory at every commit. This means elements that appear unrelated to performance but increase the size of the tree (e.g. clients using large credentials or signature keys) may provoke an unexpected increase in the cost of group operations. 

\subsection{Comparison with other works}

Table \ref{tab:comparison} shows a comparison between our work and other works that perform an analysis of performance of CGKA. Our experimental approach allows us to measure CPU time in generating and processing messages. Our work covers topics whose impact in performance has not been studied in any other of the listed works, such as External Joins, the choice of paradigm and the importance of the Delivery Service. 

Notably, other works that analyse the theoretical performance of CGKA \cite{cost_1, cost_2, cocoa, cost_tree} focus on \textit{communication cost}, which is defined as the number of ciphertexts required to heal from corruption. Our experimental analysis serves as a complement to these theoretical bounds by providing a real-world reference of the size of the exchanged messages as well as the cost of generating and processing them. Furthermore, a theoretical approach does not adequately capture the cost of maintaining and storing the Ratchet Tree, which our analysis shows to be much higher than that of cryptographic functions.

\begin{table*}[t]
\centering
\resizebox{\textwidth}{!}{
\begin{tabular}{rccccccccc}
\toprule
Work                   & Protocol     & Perspective  & Group Size     & Time   & Bandwidth  & Tree State & Paradigm & Join   & DS     \\ \midrule
Alwen et al. \cite{saik}           & Variant (SAIK) & Theoretical  & N/A            & \xmark & \cmark     & \cmark     & \xmark   & \xmark & \cmark \\ 
Auerbach et al. \cite{cost_1} & CGKA         & Theoretical  & N/A            & \xmark & \cmark     & \cmark     & \xmark   & \xmark & \xmark \\ 
Anastos et al. \cite{cost_2} & CGKA         & Theoretical  & N/A            & \xmark & \cmark     & \cmark     & \xmark   & \xmark & \xmark \\ 
Weidner et al. \cite{dec_ack}        & Variant (DGKA) & Experimental & $\approx$ 128  & \cmark & \cmark     & \xmark     & \xmark   & \xmark & \xmark \\
Chevalier et al. \cite{bonsai}          & MLS         & Experimental & $\approx$ 1,000 & \xmark       & \xmark    & \cmark    & \xmark & \xmark   & \xmark \\ 
Balbás et al. \cite{a_cgka}         & Variant (A-CGKA) & Experimental & $\approx$ 128  & \cmark & \xmark     & \xmark     & \xmark   & \xmark & \xmark \\
Paillat et al. \cite{discreet}       & MLS          & Theoretical  & N/A            & \xmark & \cmark     & \xmark     & \xmark   & \xmark & \cmark \\ 
Ours                  & MLS          & Experimental & $\approx$ 10,000 & \cmark & \cmark     & \cmark     & \cmark   & \cmark & \cmark \\ \hline
    \end{tabular}}
\caption{Comparison between works that analyse CGKA performance and ours. The columns indicate if the works consider the following topics. \textit{Time}: update generation/processing time. \textit{Bandwidth}: message size and communication cost. \textit{Tree State}: Influence of the current state of the Ratchet Tree. \textit{Paradigm}: influence of chosen paradigm. \textit{Join}: External Joins. \textit{DS}: influence of the Delivery Service.}
\label{tab:comparison}
\end{table*}

When compared with other experimental evaluations such as \cite{dec_ack, bonsai, a_cgka}, the increased depth of our analysis allows us to extract more robust conclusions. Indeed, the increments in computing cost at group sizes that are powers of 2 are not captured by works that limit this metric to 100 or 1,000 members. Our analysis also shows that cryptographic operations represent a very small percentage of experimental cost. Thus, claims about a CGKA variant's efficiency must take into account the disproportionate cost of storage and tree management that may be unrelated to its contribution.

\subsection{Limitations}

In this work, we employ an ideal Delivery Service that is readily accessible to all members and that ensures that messages are delivered in order. Thus, our analysis does not capture real-world inconveniences that can be introduced by the network such as limited bandwidth, packet losses or out-of-order delivery. MLS has also sparked interest for its application in distributed environments \cite{decaf, aa_cgka, dmls}; replacing the centralised Delivery Service employed in our testbed with a distributed one could be of interest to analyse its implications on performance.

Our analysis is limited to 10,000 users because of the significant cost of the execution of the experiments. For reference, a single commit in a 10,000-member group takes 0.1 seconds to be processed: if executed sequentially, it takes more than 15 minutes for all members to process it. In order to measure performance in larger groups, a stronger effort in parallelism is required. Alternatively, larger analysis could also ignore memory storage in order to remove the largest computational cost. 

The results obtained in this work show the relevance of handling the state of the Ratchet Tree. Thus, future work could focus on studying alternatives that minimise its size, such as compression. Similarly, the performance of clients that do not possess a copy of the Ratchet Tree ---such as the recently proposed Partial MLS~\cite{mls_partial_draft}--- could be compared to our results.

\section{Conclusion and Future Work}
\label{sec:conclusion}

In this work we have analysed the performance of MLS in an empirical setting. We analyse the protocol under different parameters such as Delivery Service, chosen paradigm or state of the Ratchet Tree. Our results show that theoretical claims of logarithmic complexity in communication cost do not manifest in practice. Our work also demonstrates the relevance of practical considerations in the performance of MLS. We have found that the most significant time costs are message distribution through the Delivery Service and the storage of the Ratchet tree, which are not taken into account in theoretical analysis. 

We also publish our execution environment as open-source, which includes a configurable emulated MLS client and two different Delivery Services. Our testbed can be employed by other researchers to perform in-depth studies of the behaviour of MLS in different environments, and can serve as a baseline for the development of more complex experiments. Our implementation can be easily adapted to execute other CGKA variants, such that their performance can be systematically compared to that of MLS. Future work should focus on implementing a more complex Delivery Service that can accurately represent real-world inconveniences and optimising the emulated client's memory usage and parallelism to allow for larger experiments.    

\section*{Acknowledgements}

The work is funded by the Plan Complementario de Comunicaciones Cuánticas, Spanish Ministry of Science and Innovation(MICINN), Plan de Recuperación NextGeneration, European Union (PRTR-C17.I1, CITIC Ref.305.2022), and Regional Government of Galicia (Agencia Gallega de Innovación, GAIN, CITIC Ref. 306.2022). D.S. acknowledges support from Xunta de Galicia and the European Union (European Social Fund - ESF) scholarship [ED481A-2023-219]. We also acknowledge support from the Xunta de Galicia and the European Union (FEDER Galicia 2021-2027 Program) Ref. ED431B 2024/21, CITIC ED431G 2023/0.

\bibliographystyle{elsarticle-num} 
\bibliography{main}

\end{document}